\documentclass[conference]{IEEEtran}
\IEEEoverridecommandlockouts
\usepackage{cite}
\usepackage{amsmath,amssymb,amsfonts}
\usepackage{algorithmic}
\usepackage{graphicx}
\usepackage{textcomp}
\usepackage{xcolor}
\usepackage{balance}
\usepackage{hyperref}
\usepackage{todonotes}
\usepackage{booktabs}
\usepackage{siunitx}
\usepackage{multirow}
\usepackage{tcolorbox}

\def\BibTeX{{\rm B\kern-.05em{\sc i\kern-.025em b}\kern-.08em
    T\kern-.1667em\lower.7ex\hbox{E}\kern-.125emX}}
    
\begin{document}
\title{Collaborative Design and Planning \\ of Software Architecture Changes \\ via Software City Visualization
}

%\title{Collaborative Planning of Refactorings \\ Using Software Cities}
%\title{Using Software Cities for \\ the Joint Planning of Refactorings}
%\title{Enhancing Refactoring Planning \\ through Collaborative Software Cities}
%\title{Joint Refactoring Planning with Software Cities}

%\author{\IEEEauthorblockN{Alexander Krause-Glau}
%\IEEEauthorblockA{\textit{Department of Computer Science} \\
%\textit{Kiel University}\\
%Kiel, Germany \\
%alexander@krause-glau.de}
%\and
%\IEEEauthorblockN{Malte Hansen}
%\IEEEauthorblockA{\textit{Department of Computer Science} \\
%\textit{Kiel University}\\
%Kiel, Germany \\
%malte.hansen@email.uni-kiel.de}
%\and
%\IEEEauthorblockN{Wilhelm Hasselbring}
%\IEEEauthorblockA{\textit{Department of Computer Science} \\
%\textit{Kiel University}\\
%Kiel, Germany \\
%hasselbring@email.uni-kiel.de}
%}

\author{
	\IEEEauthorblockN{
		Alexander Krause-Glau\IEEEauthorrefmark{1},
		Malte Hansen\IEEEauthorrefmark{2} and
		Wilhelm Hasselbring\IEEEauthorrefmark{3}
	}
	\IEEEauthorblockA{
		Department of Computer Science,
		Kiel University\\
		Kiel, Germany\\
		Email: \IEEEauthorrefmark{1}alexander@krause-glau.de,
		\IEEEauthorrefmark{2}malte.hansen@email.uni-kiel.de,
		\IEEEauthorrefmark{3}hasselbring@email.uni-kiel.de
	}
}

\maketitle

\begin{abstract}
%In professional software development, teams adhere to a structured methodology for the joint planning of software modifications.
%In this context, 
Developers usually use diagrams and source code to jointly discuss and plan software architecture changes.
%These modifications include, for example, restructurings, maintenance operations, and new features.
%However, structural diagrams provide static views and often rely on specific semantics and custom styles that developers need to understand and memorize.
%Conversely, pure source code fails to easily highlight (distributed) software communication and lacks a comprehensive high-level perspective.
With this poster, we present our on-going work on a novel approach that enables developers to collaboratively use software city visualization to design and plan software architecture changes.
\end{abstract}

\begin{IEEEkeywords}
software visualization, program comprehension, dynamic analysis, software architecture changes
\end{IEEEkeywords}

%\section{Introduction}\label{sec:introduction}
\begin{figure*}
	\includegraphics[width=\linewidth]{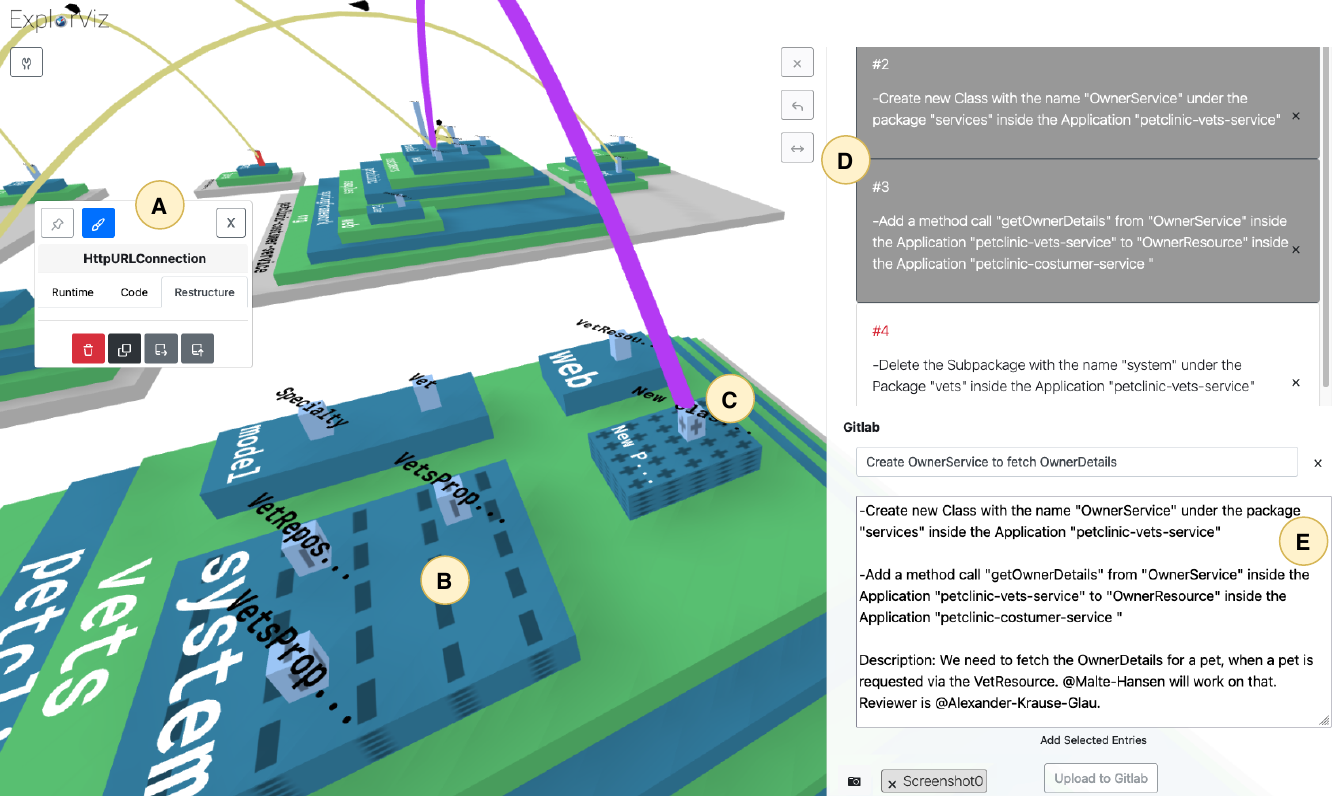}
	\caption{Screenshot of our prototype using an exemplary software city visualization encompassing multiple applications. When a user hovers the mouse over an entity, a popup emerges, displaying buttons for modifying the entity. Textures are utilized to highlight modified entities, while non-default colors indicate the entities selected by collaborators. The right sidebar features a changelog of modifications, allowing users to insert selected entries into the issue form. Additionally, users have the option to append comments or screenshots prior to uploading the issue to GitLab.}
	\label{fig:restructure-software-city}
\end{figure*}
\paragraph*{\textbf{Introduction}} Software systems are continuously changed to meet and improve both functional and non-functional requirements~\cite{williams2010CharacterizingSoftwareArchitectureChanges, latozaStudyMentalModels}.
These changes range from initial development concerns to subsequent refactorings~\cite{golubevIntelliJStudyRefactorings} or potential modernization~\cite{khadka2014WhyModernization} and maintenance~\cite{lientz1978SoftwareMaintenace}.
Due to the size, scope, and complexity, effective and efficient planning of software architecture changes (SAC) often requires a deep understanding of the existing structures and behavior of the software systems.
Although collaboration is helpful for the underlying task of program comprehension~\cite{maalej2014ProgramComprehension}, the actual designing and planning of SAC still require additional tools.
In this context, developers usually rely on familiar means such as diagrams, documentation, experience, conversation, and the actual source code~\cite{xia2018, maalej2014ProgramComprehension}.
However, diagrams and documentation tend to present a static, notationally different or unknown~\cite{purchase2003UMLNotationVariationBook}, and most often non-interactive view of software systems, i.e., they only show one level of detail~\cite{fernandez2014lodUMLComprehension}.
Source code, on the other hand, does not easily convey high-level views or (distributed) communication~\cite{cornelissen2011ControlledExperimentProgramComprehension, fittkau2015TraceControlledExperiment}.
In addition, gaining and maintaining a necessary understanding of the software structure in this context, such as through program comprehension, presents a time-consuming challenge~\cite{xia2018, tiarks2011}.
This is especially the case for new developers and also generally in the context of distributed software systems with frequent changes of the developers' mental models~\cite{tiarks2012}.
With this poster, we present our on-going work of a novel approach that enables developers to collaboratively use software cities for designing and planning SAC.
To achieve that, we extend ExplorViz~\cite{VISSOFT2013,fittkau2017IST,hasselbring2020} and its collaboration mode~\cite{krauseglau2022vissoft} with modifiable software cities.
\paragraph*{\textbf{Approach}}Figure~\ref{fig:restructure-software-city} depicts a screenshot of our current implementation.
Here, we see multiple software cities visualizing an extended version of the distributed Spring PetClinic.\footnote{\url{https://github.com/spring-petclinic/spring-petclinic-microservices}}
The foundation for this visualization is a dynamic analysis of the PetClinic's runtime behavior.
Owing to ExplorViz's architecture~\cite{krauseglau2022ic2e}, this dynamic analysis can be executed within continuous integration pipelines.
This enables users to automatically update the visual representation of the software system, specifically the software city visualizations, in contrast to most diagram tools.

Regarding the design of SAC, users primarily interact with ExplorViz through extended popups.
These appear when an entity, e.g., a class depicted as blue building, is hovered over (Figure~\ref{fig:restructure-software-city}-A).
The popups provide a set of options for manipulating the software city entities.
Each popup is context-sensitive, meaning that the available buttons and actions are tailored to the specific type of entity being hovered over.
Modifications are highlighted via textures, e.g., plus signs, instead of color strategies to reduce noise and facilitate comprehension (Figure~\ref{fig:restructure-software-city}-B and Figure~\ref{fig:restructure-software-city}-C).
Each modification is recorded in a changelog window as shown in Figure~\ref{fig:restructure-software-city}-D.
Here, users have an overview of all modifications.
Clicking on an entry's number highlights the corresponding modification within the visualization.
This feature is intended to mitigate potential scalability issues related to the visualization.
Furthermore, the changelog provides users with the capability to undo each modification.

Eventually, users can generate GitLab issues directly from within the software visualization tool using the issue form in Figure~\ref{fig:restructure-software-city}-E.
In this process, users can select specific changelog entries to include in an issue.
Additionally, the issue can be named and supplemented with additional comments and screenshots.
In this instance, we also utilized GitLab usernames, which highlight the issue for the respective users in GitLab.
%\input{chapters/related-work.tex}

%\section*{Acknowledgment}
\paragraph*{\textbf{Acknowledgment}} The authors would like to thank Arash Giv for his contributions to the implementation.

%\clearpage
\providecommand{\doi}[1]{DOI: \href{https://doi.org/#1}{#1}}
\bibliographystyle{myIEEEtran}
%\interlinepenalty=10000
\balance
\bibliography{explorviz-vissoft-2024-restructure}

\end{document}